\documentclass[10pt,journal]{IEEEtran}
\IEEEoverridecommandlockouts
\usepackage{amsmath,amssymb,amsfonts}
\usepackage{algorithmic}
\usepackage{algorithm}
\usepackage{array}
\usepackage{textcomp}
\usepackage{stfloats}
\usepackage{url}
\usepackage{verbatim}
\usepackage{graphicx}
\usepackage{cite}
\usepackage{subcaption}
\usepackage{bm}
\usepackage{makecell}
\usepackage{diagbox}
\usepackage{multirow}
\usepackage{pifont}
\usepackage{soul,color}
\usepackage{xurl}
\usepackage{enumitem}
\usepackage{booktabs}

\begin{document}

\title{Capacity-Optimized Pre-Equalizer Design \\ for Visible Light Communication Systems
}

\author{Runxin Zhang, Yulin Shao, Jian Xiong, Lu Lu, Murat Uysal
\thanks{R. Zhang and L. Lu are with the Key Laboratory of Space Utilization, Technology and Engineering Center for Space Utilization, Chinese Academy of Sciences, Beijing, 100094, China, and with the University of Chinese Academy of Sciences, China.
R. Zhang is also with the Engineering Division, New
York University Abu Dhabi, UAE (emails: zhangrunxin20@mails.ucas.ac.cn, lulu@csu.ac.cn).

Y. Shao is with the Department of Electrical and Electronic Engineering, University of Hong Kong, Hong Kong S.A.R. (e-mail: ylshao@hku.hk).

J. Xiong is with the Hangzhou Institute for Advanced Study, University of Chinese Academy of Sciences, Hangzhou, 310024, China (e-mail: xiongjian@ucas.ac.cn).

M. Uysal is with the Engineering Division, New
York University Abu Dhabi, UAE (e-mail: murat.uysal@nyu.edu).
}
}

\maketitle

\begin{abstract}
Since commercial LEDs are primarily designed for illumination rather than data transmission, their modulation bandwidth is inherently limited to a few MHz. This becomes a major bottleneck in the implementation of visible light communication (VLC) systems necessiating the design of pre-equalizers. While state-of-the-art equalizer designs primarily focus on the data rate increasing through bandwidth expansion, they often overlook the accompanying degradation in signal-to-noise ratio (SNR). Achieving effective bandwidth extension without introducing excessive SNR penalties remains a significant challenge, since the channel capacity is a non-linear function of both parameters. In this paper, we present a fundamental analysis of how the parameters of the LED and pre-equalization circuits influence the channel capacity in intensity modulation and direct detection (IMDD)-based VLC systems. We derive a closed-form expression for channel capacity model that is an explicitly function of analog pre-equalizer circuit parameters. Building upon the derived capacity expression, we propose a systematic design methodology for analog pre-equalizers that effectively balances bandwidth and SNR, thereby maximizing the overall channel capacity across a wide range of channel attenuations. We present extensive numerical results to validate the effectiveness of the proposed design and demonstrate the improvements over conventional bandwidth-optimized pre-equalizer designs.
\end{abstract}

\begin{IEEEkeywords}
Visible light communication, Analog equalizer, IMDD channel capacity
\end{IEEEkeywords}

\section{Introduction}
\label{SecI}
The exponential growth of mobile data traffic, coupled with the increasing quality-of-service demands of emerging applications, has pushed conventional radio-frequency (RF) networks to their fundamental capacity limits \cite{chi2020visible,chowdhury20206g,alsabah20216g}, resulting in severe spectrum congestion. This pressing challenge has spurred the exploration of alternative wireless technologies. Among these, the unlicensed optical spectrum in the visible light band has emerged as a promising complement to RF systems. Visible light communication (VLC) transmits data by modulating the intensity of light-emitting diodes (LEDs) and receives it via photodiode (PD)-based detectors. VLC offers three key advantages: license-free operation, inherent physical-layer security, and the dual functionality of providing both illumination and data transmission \cite{chow2024recent,matheus2019visible,pathak2015visible}. Recognizing its potential, the IEEE introduced the 802.11bb standard in 2023, formally integrating VLC into the existing wireless ecosystem through the definition of physical layer (PHY) and medium access control layer (MAC) protocols.

While the optical spectrum is large, the bottleneck in the implementation of VLC systems comes from the limited electrical bandwidth of LEDs. Since commercial LEDs are primarily designed for illumination rather than data transmission, their modulation bandwidth is inherently limited to a few MHz due to carrier recombination dynamics and parasitic effects within the semiconductor structure \cite{rashidi2017differential}. To address this limitation, equalization techniques have been explored in the literature \cite{kisacik2022130,zhang2024design,zhang2019over,li2020should,kassem2021use,chen2021digital,li2023performance}. Based on the location of their deployment, they are categorized as pre-equalization at the transmitter and post-equalization at the receiver \cite{wang202175}. Post-equalization employs adaptive signal processing at the receiver to dynamically compensate for channel-induced distortions and frequency-selective fading. It offers strong adaptability to time-varying channel conditions, enabling real-time adjustments to abrupt fluctuations through iterative algorithms \cite{li2023performance,zhao2021low,zhang202416qam}. While post-equalization simplifies transmitter design, it assumes that the received signal maintains sufficient fidelity for effective compensation - a condition that is difficult to achieve in bandwidth-limited optical links. Pre-equalization is therefore critical to fully harness the modulation bandwidth potential of VLC systems where the bottleneck stems from the limited modulation bandwidth of LEDs. It effectively mitigates bandwidth constraints while also reducing receiver complexity and processing latency.

Pre-equalization can be further divided into analog pre-equalization and digital pre-equalization \cite{zhang2024design,li2020should,yang2023visible}. Digital pre-equalization offers the advantage of precise control over the equalization process, allowing it to correct for non-linearities and frequency-dependent impairments inherent in the optical channel. Compared to digital pre-equalization, analog pre-equalization does not require high-speed digital signal processing or high-precision digital-to-analog converters. Instead, it provides a simpler solution by directly modifying the transmitted signal in the analog domain before it is applied to the LED \cite{zhang2024design}.

Most existing studies on equalization for VLC primarily focus on increasing data rates through bandwidth enhancement \cite{kisacik2022130,zhang2024design,zhang2019over}, while only a limited number of works have considered that excessive bandwidth extension can lead to undesirable signal power attenuation. A comparison of these studies is provided in Tab.~\ref{tab:review}. For example, the authors of \cite{li2020should} and \cite{chen2021digital} recognized that analog equalization can significantly attenuate signal power. The authors of \cite{li2020should} argued that equalizers may be unnecessary in systems employing bit loading, while the authors of \cite{chen2021digital} proposed a distributed digital pre-equalization technique, which can be viewed as a combination of digital pre-equalization and bit loading across multiple sub-bands, thereby achieving a higher achievable rate. The authors of \cite{kassem2021use} addressed the issue of signal attenuation by implementing a negative capacitance circuit using transistors, which directly compensates for the intrinsic capacitance of the LED, thereby extending the bandwidth without degrading signal power. Inspired by the gain-bandwidth product metric used in photodiodes, the authors of \cite{chen2023oma} introduced the optical modulation amplitude and bandwidth product (OBP) as a new merit to assess the channel capacity potential of different LEDs. However, they did not offer specific guidance on how to fully exploit each LED’s capacity potential.

\begin{table*}[!tb]
\centering
\caption{Comparison of Equalization Techniques.}
\begin{tabular}{cccccc}
     \toprule
     \textbf{Refs.}   & \textbf{Performance Metric} & \makecell{\textbf{Trade-off} \\ \textbf{Analysis}} & \makecell{\textbf{Equalization Type} \\ \textbf{(Pre/Post)}} & \makecell{\textbf{Equalization Type} \\ \textbf{(Analog/Digital)}} & \textbf{Equalization Approach} \\
     \midrule
     \cite{kisacik2022130} & Bandwidth  & & Post & Analog & \makecell[l]{ Switchable-capacitor equalizer } \\
     \cite{zhang2024design} & Bandwidth & & Pre & Analog & \makecell[l]{RCL-based passive equalizer} \\
     \cite{li2020should} & Channel capacity  & \ding{51} & Pre & Digital & \makecell[l]{Bit-loading-based multicarrier modulation}\\
     \cite{kassem2021use} & Bandwidth and attenuation & & Pre & Analog & \makecell[l]{Transistor-based negative capacitance circuit} \\
     \cite{chen2021digital} & Data rate & \ding{51} & Pre & Digital & \makecell[l]{Distributed equalization} \\
     \cite{li2023performance} & Data rate & & Post & Digital & \makecell[l]{GRU neural network-based equalizer} \\
     \cite{chen2023oma} & OBP & \ding{51} & Pre & Analog & \makecell[l]{RCL-based passive equalizer} \\
     \textbf{This paper} & \textbf{Channel capacity}  & \textbf{\ding{51}} & \textbf{Pre} & Analog & \makecell[l]{ \textbf{Capacity-optimized equalizer design}} \\
     \bottomrule
\end{tabular}
\label{tab:review}
\end{table*}

\begin{figure}
    \centering
    \includegraphics[width=0.8\linewidth]{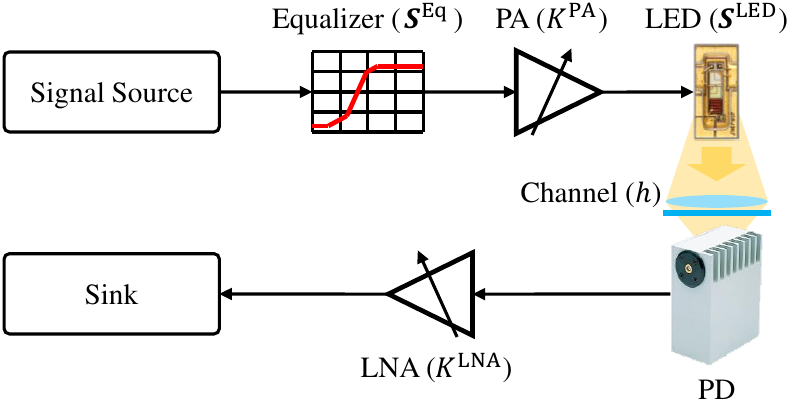}
    \caption{Schematic of the optical wireless transmission link. The system is modeled as a cascaded two-port network, where each module contributes to the overall transmission characteristics.}
    \label{fig:optical_link}
\end{figure}

While optimized modulation schemes in the digital domain can enhance the achievable rate, optimization of channel capacity through equalization remains an open problem. This is particularly important in achieving effective bandwidth extension without incurring excessive signal-to-noise ratio (SNR) penalties. To bridge this gap, we present a fundamental analysis of how the parameters of the LED and equalization circuits influence the channel capacity in intensity modulation and direct detection (IMDD)-based VLC systems. Then, we formulate an optimization problem to maximize the channel capacity whose solution yields the optimal equalization parameters. Our study further derives an explicit mathematical expression linking the optimal solution to the channel attenuation. This provides a general guideline for rapid computation and system design under varying channel attenuation conditions, thereby enhancing the applicability and generalizability of the proposed approach. The main contributions of our paper are summarized as follows:
\begin{itemize}
    \item Within the IMDD framework, we derive a closed-form expression for channel capacity model that is an explicitly function of analog equalizer circuit parameters.
    \item Building upon the derived capacity expression, we propose a systematic design methodology for analog equalizers that effectively balances bandwidth and SNR, thereby maximizing the overall channel capacity across a wide range of channel attenuation values.
    \item We conduct an extensive simulation study to validate the effectiveness of the proposed design. The derived closed-form expressions for the optimal pole locations achieve a normalized mean square error (NMSE) below 3\%, while the closed-form solutions corresponding to maximum capacity maintain errors below 5\% across varying channel attenuation.
\end{itemize}

The remainder of this paper is organized as follows. Section \ref{SecII} presents the system model, detailing the equivalent circuit models for LED and equalization circuits. Section \ref{SecIII} derives the channel capacity as an explicit function of system parameters. Section \ref{SecIV} formulates the capacity-optimized equalizer design problem and analytically solves it to obtain optimized equalizer parameters. Section \ref{SecV} verifies the proposed design through simulation results. Section \ref{SecVI} concludes this paper.

\section{System Model}
\label{SecII}
We consider a VLC system as illustrated in Fig.~\ref{fig:optical_link}, which consists of a commercial monochromatic LED and a broadband PD. The transmitter and receiver are equipped with a power amplifier (PA) and a low-noise amplifier (LNA), respectively, with corresponding gains $K^\text{PA}$ and $K^\text{LNA}$. An analog equalizer is integrated into the transmitter to compensate for bandwidth limitations. All ports of active components, including the signal source and amplifiers, are assumed to operate with a standard impedance of $R_g$  which is typically chosen as 50 $\Omega$ in practice implementation. The channel attenuation of the VLC propagation channel is denoted as $h \in \mathbb{R}$.

\begin{figure}[t]
    \centering
    \includegraphics[width=.75\linewidth]{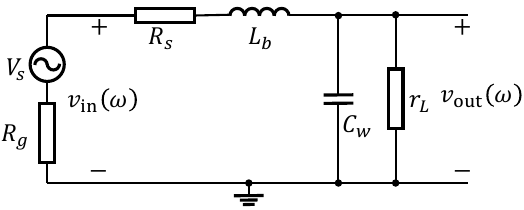}
    \caption{Second-order equivalent circuit for an LED.}
    \label{f:LED_2nd_order}
\end{figure}

LED is a semiconductor device, and can be divided into two key components: the cladding layer and the multiple quantum well (MQW) layer. In our study, we adopt the second-order equivalent circuit to characterize the LED, as shown in Fig.~\ref{f:LED_2nd_order}. We denote the frequency response characteristics of the second-order equivalent circuit by
\begin{eqnarray} \label{e:S_LED}
    \hspace{-0.5cm} &&\bm{S}^{\text{LED}}     \triangleq     
    \begin{bmatrix}
        S_{11}^{\text{LED}} &     S_{12}^{\text{LED}} \\
        S_{21}^{\text{LED}} &     S_{11}^{\text{LED}} 
    \end{bmatrix},
\end{eqnarray}
where $S_{11}^{\text{LED}}$ and $S_{22}^{\text{LED}}$ are the input and output reflection coefficients, indicating how much of the signal is reflected back at each port; $S_{21}^{\text{LED}}$ and $S_{12}^{\text{LED}}$ describe the forward and reverse transmission coefficients, measuring how much power is transmitted from one port to the other.

\begin{figure}
    \centering
    \includegraphics[width=0.5\linewidth]{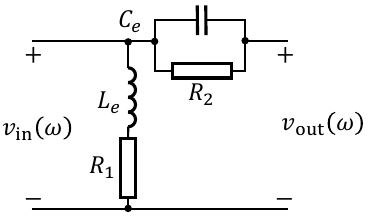}
    \caption{The architecture of the general second-order equalization circuit.}
    \label{fig:equalizer}
\end{figure}

To overcome the bandwidth limitations caused by the intrinsic electrical characteristics of the LED, we employ a general second-order equalizer \cite{zhang2024design}, as illustrated in Fig.~\ref{fig:equalizer}, to enhance the effective bandwidth of the VLC system under consideration. The scattering matrix of the equalizer circuit is denoted by
\begin{eqnarray} \label{e:S_Eq}
    \hspace{-0.5cm} && \bm{S}^{\text{Eq}}      \triangleq     
    \begin{bmatrix}
        S_{11}^{\text{Eq}} &     S_{12}^{\text{Eq}} \\
        S_{21}^{\text{Eq}} &     S_{11}^{\text{Eq}} 
    \end{bmatrix}.
\end{eqnarray}
Without loss of generality, we assume that the PA and LNA provide linear amplification of the input signal with perfect impedance matching. The corresponding scattering matrix of the PA and LNA are defined, respectively, as
\begin{eqnarray} 
    \bm{S}^{\text{PA}}      \triangleq     
    \begin{bmatrix}
        0 &     0 \\
        K^{\text{PA}} &     0
    \end{bmatrix}
    , \quad \text{and} \quad 
    \bm{S}^{\text{LNA}}      \triangleq     
    \begin{bmatrix}
        0 &     0 \\
        K^{\text{LNA}} &     0
    \end{bmatrix}.
\end{eqnarray}

\section{IMDD Channel Capacity for the Equalized VLC System}
\label{SecIII}
The channel capacity for an IMDD system, where signals are strictly positive, is given by \cite{chaaban2021capacity} 
\begin{eqnarray} \label{e:C}
    \mathcal{C} = \frac{1}{2} \mathcal{B} \log_2{\left( \frac{(K^{\text{C}})^2 \mu^2}{8 \pi e \mathcal{B} N_0} + 1 \right) }.
\end{eqnarray}
where $\mathcal{B}$ is  the 3dB-bandwidth of the equalized system and $K^{\text{C}}$  denotes the effective (end-to-end) channel coefficient of the equalized system that incorporates the combined effect of transmitter/receiver front-ends and the propagation channel. In the following, building on the scattering matrices of each module introduced in the previous section, we first present the frequency response function of the VLC link formed by the cascade of these modules. Then, we obtain the closed form expressions of end-to-end channel coefficient term and 3 dB bandwidth in terms of the system parameters. 

For each scattering matrix $\bm{S}$, we define a transfer matrix \cite{egan2004practical}
\begin{eqnarray} \label{e:T_matrix}
    \bm{T} \triangleq \begin{bmatrix}
        T_{11} & T_{12} \\
        T_{21} & T_{22}
    \end{bmatrix}
    = \begin{bmatrix}
        \frac{1}{S_{21}} & - \frac{S_{22}}{S_{21}} \\
        \frac{S_{11}}{S_{21}} & \frac{S_{12} S_{21} - S_{11} S_{22}}{S_{21}}
    \end{bmatrix}.
\end{eqnarray}

The transfer matrices corresponding to the equalization circuit, PA, and LED are denoted by $\bm{T}^{\text{Eq}}$, $\bm{T}^{\text{PA}}$, and $\bm{T}^{\text{LED}}$, respectively. Given the cascaded configuration of these components, the overall transfer matrix of the system can be expressed as
\begin{eqnarray} \label{e:T^C}
    \bm{T}^{\text{C}} = \bm{T}^{\text{Eq}} \bm{T}^{\text{PA}} \bm{T}^{\text{LED}}
    = \begin{bmatrix}
        \frac{1}{S^{\text{Eq}}_{21}} \frac{1}{K^{\text{PA}}} \frac{1}{S^{\text{LED}}_{21}} & - \frac{S^{\text{LED}}_{22}}{S^{\text{LED}}_{21}} \frac{1}{S^{\text{Eq}}_{21}} \frac{1}{K^{\text{PA}}} \\
        \frac{S^{\text{Eq}}_{11}}{S^{\text{Eq}}_{21}} \frac{1}{S^{\text{LED}}_{21}} \frac{1}{K^{\text{PA}}} & - \frac{S^{\text{LED}}_{22}}{S^{\text{LED}}_{21}} \frac{S^{\text{Eq}}_{11}}{S^{\text{Eq}}_{21}} \frac{1}{K^{\text{PA}}}
    \end{bmatrix}.
\end{eqnarray}

Based on the circuit structure in Fig.~\ref{f:LED_2nd_order}, the forward transmission coefficient of the LED can be derived as
\begin{eqnarray} \label{e:S21_LED}
    S_{21}^{\text{LED}} & = & \frac{\frac{2 r_L}{\left( R_s + R_g \right) \left( r_L + 1 \right)}}{\left(  1 + \frac{j \omega L_b}{R_s + R_g}  \right)  \left( 1 + \frac{j \omega C_w r_L}{r_L +1}  \right)} \notag\\
    & \triangleq & \frac{K^{\text{LED}}}{\left( 1 + \frac{j \omega}{\omega^{\text{LED}}_{p1}} \right)  \left( 1+ \frac{j \omega}{\omega^{\text{LED}}_{p2}} \right)},
\end{eqnarray}
where $j$ represents the imaginary unit, $\omega$ is the angular frequency, and $\frac{1}{j \omega C}$, $j \omega L$ and $R$ correspond to the impedances of the capacitor, inductance and resistor, respectively. $C_w$ is defined as the capacitance associated with the confined carriers in the MQW; $r_L$ represents the resistances due to carriers in the cladding and MQW regions; $L_b$ is the inductance associated with unconfined carriers in the cladding region and space charge; and $R_s$ is the series resistance. We further define$\omega_{p1}^{\text{LED}} \triangleq \frac{r_L + 1}{C_w r_L} = 2 \pi f_{p1}^{\text{LED}}$ and $\omega_{p2}^{\text{LED}} \triangleq \frac{R_s + R_g}{L_b} = 2 \pi f_{p2}^{\text{LED}}$ as two poles with $\omega_{p1}^{\text{LED}} \leq \omega_{p2}^{\text{LED}}$, and $K^{\text{LED}} \triangleq \frac{2 r_L}{(R_s + R_g) (r_L + 1)} = \frac{2}{4 \pi^2 C_w L_b} \frac{1}{f_{p1}^{\text{LED}} f_{p2}^{\text{LED}}}$ being LED response gain.

On the other hand, the scattering parameter of the equalizer can be derived from the circuit structure in Fig.~\ref{fig:equalizer} as
\begin{eqnarray} \label{e:S21_Eq}
    S_{21}^{\text{Eq}} & = & \frac{ \frac{4  R_{g}  R_{1} }{ 2  R_{1}  +  R_{g} }  \frac{1}{2  R_{g}  +  R_{2}}  \left[  \frac{j \omega}{ \frac{R_{1} }{L_{e}}  +  1}  \right]  \left[  \frac{j \omega}{ \frac{ 1 }{ C_{e}  R_{2}}  +  1}  \right]}{ \left[  \frac{j \omega}{ \frac{ 2  R_{1}  +  R_{g} }{ 2 L_e}  +  1}  \right]  \left[  \frac{j \omega}{ \frac{ 2  R_{g}  +  R_{2} }{ 2  R_{g}  C_{e}  R_{2}}  +  1}  \right] } \notag\\
    & \triangleq & \frac{K^{\text{Eq}} \left(1 + \frac{j \omega}{\omega^{\text{Eq}}_{z1}} \right) \left( 1 + \frac{j \omega}{\omega^{\text{Eq}}_{z2}} \right)}{\left(1 + \frac{j \omega}{\omega^{\text{Eq}}_{p1}} \right) \left( 1 + \frac{j \omega}{\omega^{\text{Eq}}_{p2}} \right)},
\end{eqnarray}
where the components $R_1$, $L_e$, $R_2$, and $C_e$ are parameters of the equalizer circuit, as shown in Fig.~\ref{fig:equalizer}. $\omega_{z1}^{\text{Eq}} \triangleq \frac{R_1}{L_e} = 2 \pi f_{z1}^{\text{Eq}}$ and $\omega_{z2}^{\text{Eq}} \triangleq \frac{1}{C_e R_2} = 2 \pi f_{z2}^{\text{Eq}}$ are two zeros with $\omega_{z1}^{\text{Eq}} \leq \omega_{z2}^{\text{Eq}}$, 
$\omega_{p1}^{\text{Eq}} \triangleq \frac{2R_1+R_g}{2L_e} = 2 \pi f_{p1}^{\text{Eq}} > 2 \pi f_{z1}^{\text{Eq}}$ and $\omega_{p2}^{\text{Eq}} \triangleq \frac{2 R_g + R_2}{2 R_g C_e R_2} = 2 \pi f_{p2}^{\text{Eq}} > 2 \pi f_{z2}^{\text{Eq}}$ are two poles, 
and $K^{\text{Eq}} \triangleq \frac{4R_g R_1}{2R_1+R_g} \frac{1}{2R_g + R_2} = \frac{\omega_{z1}^{\text{Eq}}}{\omega_{p1}^{\text{Eq}}} \frac{\omega_{z2}^\text{Eq}}{\omega_{p2}^\text{Eq}} = \frac{f_{z1}^\text{Eq}}{f_{p1}^\text{Eq}} \frac{f_{z2}^\text{Eq}}{f_{p2}^\text{Eq}}$ is response gain of the equalization circuit.

Combining \eqref{e:T_matrix}-\eqref{e:S21_Eq}, the overall system frequency response function can be 
\begin{eqnarray} \label{e:H}
    H(f) & = & \mathcal{R}_{L}  h  \mathcal{R}_{P} K^{\text{LNA}} S_{21}^{\text{C}} (2 \pi f) \notag\\ 
    & = & \frac{  \mathcal{R}_{L}  h \mathcal{R}_{P} K^{\text{LNA}}K^{\text{Eq}} K^{\text{PA}} K^{\text{LED}}}{\left(\frac{j f}{ f^{\text{Eq}}_{p1}} + 1 \right) \left(\frac{j f}{ f^{\text{Eq}}_{p2}} + 1 \right)}.
\end{eqnarray}
where $\mathcal{R}_L$ is the electro-optical conversion efficiency of the LED, $\mathcal{R}_P$ is the responsivity of the PD, and $S_{21}^{\text{C}}= \frac{1}{T_{11}^{\text{C}}}$ is the forward transmission coefficient of the cascaded transmitter. 
Given the frequency response, the 3-dB bandwidth of the VLC system can be derived as (see Appendix \ref{sec:AppA})
\begin{eqnarray} \label{e:B}
    \mathcal{B}(R_1, R_2, L_e, C_e) &=& \frac{ 1 }{ |H(f)|^2_{ \text{max} }} \int_0^\infty {|H(f)|^2 df} \notag \\
    &=& \frac{\pi}{2} \frac{f^{\text{Eq}}_{p1} f^{\text{Eq}}_{p2}}{f^{\text{Eq}}_{p1} + f^{\text{Eq}}_{p2}}.
\end{eqnarray}
The end-to-end channel coefficient is
\begin{eqnarray} \label{e:K^C}
    K^{\text{C}} \hspace{-0.6 cm}&&(R_1, R_2, L_e, C_e)  =    \mathcal{R}_{L}  h \mathcal{R}_{P} K^{\text{LNA}}  K^{\text{Eq}} K^{\text{PA}}  K^{\text{LED}} \notag\\
    \hspace{-0.2 cm}&& \overset{(a)}{=}  \mathcal{R}_{L}  h \mathcal{R}_{P} K^{\text{LNA}}  K^{\text{PA}} \frac{ f_{z1}^{\text{Eq}} f_{z2}^{\text{Eq}} }{f_{p1}^{\text{Eq}} f_{p2}^{\text{Eq}}} \frac{2}{C_w L_b}\frac{1}{4 \pi ^2  f_{p1}^{\text{LED}} f_{p2}^{\text{LED}}}  \notag\\
    \hspace{-0.2 cm}&& \overset{(b)}{=}  \frac{ \mathcal{R}_{L}  h \mathcal{R}_{P} K^{\text{LNA}}  K^{\text{PA}} }{2 \pi ^2 C_w L_b f_{p1}^{\text{Eq}} f_{p2}^{\text{Eq}}} ,
\end{eqnarray}
where (a) follows from the explicit expressions of  $K^{\text{LED}}$ and  $K^{\text{Eq}}$ obtained in \eqref{e:S21_LED} and \eqref{e:S21_Eq}, and (b) follows from the zeros-pole matching method \cite{zhang2024design}. Replacing \eqref{e:K^C} in \eqref{e:C}, we have 
\begin{eqnarray} \label{e:capacity}
    \hspace{-1 cm}&& \mathcal{C}(R_1, R_2, L_e, C_e) \overset{(a)}{=} \frac{\pi}{2}  \frac{f^{\text{Eq}}_{p1} f^{\text{Eq}}_{p2}}{f^{\text{Eq}}_{p1} + f^{\text{Eq}}_{p2}} \frac{1}{2}  \\
    \hspace{-1 cm}&& \cdot \log_2{ \left( \frac{\left( \mathcal{R}_{ L} h \mathcal{R}_{ P} K^{ \text{LNA}} K^{ \text{PA}} \right)^2 \mu^2}{ 4 \pi ^2 e N_0\left( 2 \pi^2 C_w L_b \right)^2} \frac{f^{\text{Eq}}_{p1} + f^{\text{Eq}}_{p2}}{\left( f^{\text{Eq}}_{p1} \right)^{ 3} \left( f^{\text{Eq}}_{p2} \right)^{ 3}} + 1  \right) }.\notag
\end{eqnarray}

It is observed from \eqref{e:B} and \eqref{e:K^C} that there is a fundamental interplay between bandwidth expansion and signal power, revealing that these two factors are intrinsically coupled through the circuit design parameters. As a result, treating bandwidth and channel coefficient as independent optimization objectives is insufficient for achieving maximum channel capacity. Instead, a joint optimization approach is essential - one that carefully balances equalizer pole placement, LED circuit characteristics, and power amplifier design to simultaneously enhance bandwidth and preserve signal integrity. This insight challenges the conventional bandwidth-optimized design paradigm in earlier works and emphasizes the necessity of a capacity-optimized equalizer design, which will be dealt with in the next section under different levels of channel attenuations.

\section{Capacity-Optimized Equalizer Design}
\label{SecIV}
In this section, we aim to optimize the derived channel capacity expression in \eqref{e:capacity} to determine the equalizer parameters. Therefore, the optimization objective is formulated as
\begin{eqnarray} 
    R_1^*, R_2^*, L_e^*, C_e^* = \arg \max_{\{R_1, R_2, L_e, C_e \}} {\mathcal{C} (R_1, R_2, L_e, C_e) }.
\end{eqnarray}

To effectively solve the channel capacity optimization problem, we begin by analyzing the impact of each design variable individually. Among these, the equalization circuit plays a crucial role in modulation bandwidth enhancement by employing the zero-pole matching method. This technique strategically positions zeros and poles in the system’s transfer function to counteract bandwidth limitations, leading to
\begin{eqnarray*} \label{e:ZPM}
    \omega_{z1}^{\text{Eq}} &=& \frac{R_1}{L_e} = \omega_{p1}^{\text{LED}} = \frac{r_L + 1}{C_w r_L},\\
    \omega_{z2}^{\text{Eq}} &=& \frac{1}{C_e R_2} = \omega_{p2}^{\text{LED}} = \frac{R_s + R_g}{L_b}.
\end{eqnarray*}
From these expressions, the optimal $L_e^*$ and $C_e^*$ can be expressed in terms of $R_1^*$ and $R_2^*$ as
\begin{eqnarray} \label{e:L_e_C_e}
    L_e^* = \frac{R_1^* C_w r_L}{r_L + 1}, \quad \text{and} \quad C_e^* = \frac{L_b }{R_2^* (R_s + R_g)}.
\end{eqnarray}
By defining $x = f_{p1}^{\text{Eq}} > f_{z1}^{\text{Eq}} = f_{p1}^{LED}$, $y = f_{p2}^{\text{Eq}} > f_{z2}^{\text{Eq}} = f_{p2}^{\text{LED}}$, and $\alpha = \frac{\left(K^{\text{PA}} \mathcal{R}_L h \mathcal{R}_P K^{\text{LNA}} \right)^2 \mu^2}{4 \pi^2 e N_0 \left(2 \pi^2 C_w L_b \right)^2} > 0$, \eqref{e:capacity} can be rewritten in a simple form as
\begin{eqnarray} \label{e:capacity_0}
    \hspace{-1 cm}&& \mathcal{C}  =  \frac{\pi}{4} \frac{x y }{x + y} \log_2{\left( \alpha \frac{x + y}{x^3 y^3 } + 1 \right)}.
\end{eqnarray}

To analyze the optimization problem, we partition x into two distinct intervals: $x \in \left[f_{p1}^{\text{LED}}, f_{p2}^{\text{LED}} \right)$ and $x \in \left[f_{p2}^{\text{LED}}, \infty \right)$.

\subsubsection{Optimal Design Under High Channel Attenuation}
We first consider the case $x \in \left[f_{p2}^{\text{LED}}, \infty \right)$ where $x$ and $y$ exhibit cyclic symmetry in $\mathcal{C}$. Taking the partial derivative with respect to $x$, we have
\begin{eqnarray*} 
    &&\hspace{-0.5 cm}  \frac{\partial \mathcal{C}}{\partial x} = \frac{\pi}{4} \frac{y(x + y) - xy}{(x + y)^2} \log_2{\left( 1 + \alpha \frac{x + y }{x^3 y^3} \right)}  \\
    &&\hspace{0.5 cm} +  \frac{\pi}{4} \frac{xy}{x + y} \left[ \frac{1}{\ln{2} \frac{1}{1 + \alpha \frac{x + y}{ x^3 y^3}} } \alpha \frac{x^3 y^3 -(x + y) 3 x^2 y^3}{x^6 y^6} \right] \notag \\
    && \hspace{-0.5 cm} =  \frac{\pi}{4} \!\frac{y^2}{(x\! +\! y)^2} \log_2{\!\!\left(\!\! 1 \!\!+\!\! \alpha \frac{x\! +\! y}{ x^3 y^3} \!\! \right)}\!\! +\! \frac{\pi}{4} \frac{\alpha}{\ln{2} } \frac{y}{x\! +\! y} \frac{-2 x - 3 y}{x^3 y^3 \!+\! \alpha (x\! +\! y)}.\notag
\end{eqnarray*}
To maximize $\mathcal{C}$, we can set $\frac{\partial \mathcal{C}}{\partial x} = 0$ and obtain
\begin{equation*} 
    \frac{\ln{2}}{\alpha} \frac{x^3 y^3 \!+\! \alpha (x\! +\! y)}{x + y} \log_2{\!\!\left(\!\! 1 \!\!+\!\! \alpha \frac{x\! +\! y}{ x^3 y^3} \!\! \right)} = \frac{2 x \!+\! 3 y}{y}.
\end{equation*}
Similarly, by setting $\frac{\partial \mathcal{C}}{\partial y} = 0$, we obtain
\begin{equation*} 
    \frac{\ln{2}}{\alpha} \frac{x^3 y^3 \!+\! \alpha (x\! +\! y)}{x + y} \log_2{\!\!\left(\!\! 1 \!\!+\!\! \alpha \frac{x\! +\! y}{ x^3 y^3} \!\! \right)} = \frac{2 y \!+\! 3 x}{x}.
\end{equation*}
The left-hand sides of the above two equations are equal. By combining these two equations, we get
\begin{equation*} 
    \frac{2 x + 3 y}{y} = \frac{2 y + 3 x}{x}.
\end{equation*}

Since $x,y>0$, $\mathcal{C}$ achieves its maximum value for $x=y$, which indicates that symmetric pole placement in the high-frequency regime maximizes the system's capacity efficiency, ensuring an optimal balance between bandwidth extension and signal power. In this context, the optimization objective can be simplified to
\begin{eqnarray} \label{e:capacity_1}
    \mathcal{C} = \frac{\pi}{4} \frac{1}{2} \log_2{\left( \frac{2 \alpha}{x^5} +1 \right)}.
\end{eqnarray}
The values of $x$ and the achievable $\mathcal{C}$ at the maximum channel capacity are provided below. First, compute the partial derivative of $\mathcal{C}$ with respect to $x$ and set $\frac{\partial \mathcal{C}}{\partial x} = 0$. Reorganizing the resulting expression, we obtain
\begin{equation*} 
    \log_2{\left(  \frac{2 \alpha}{x^5} + 1 \right)} = \frac{5}{\frac{x^5}{\alpha} + 1}.
\end{equation*}
The solution for $x$ is given by (see the proof in Appendix \ref{sec:AppB})
\begin{eqnarray} \label{e:optimal_x_1}
    x^* = \frac{(2 \alpha)^{\frac{1}{5}}}{e}
\end{eqnarray}
Once the optimal pole $x^*$ is determined, the corresponding optimal parameter values can be obtained by substituting it into the relationships given in \eqref{e:S21_Eq} and \eqref{e:L_e_C_e} which yield
\begin{eqnarray*} 
    R_1^* & = & \frac{ R_g (r_l + 1)}{4 \pi C_w r_L (2 \alpha)^{\frac{1}{5}} - 2 e (r_L + 1) }, \\
    L_e^* & = &  \frac{ e R_g C_w r_L}{4 \pi C_w r_L (2 \alpha)^{\frac{1}{5}} - 2 e (r_L + 1) }, \\
    R_2^* & = & \frac{4 \pi R_g L_b }{R_s + R_g}\frac{(2 \alpha)^{\frac{1}{5}}}{e} - 2 R_g, \\
    C_e^* & = &  \frac{ e L_b}{4 \pi R_g L_b (2 \alpha)^{\frac{1}{5}} - 2 e R_g (R_s + R_g) }.
\end{eqnarray*}
Substituting \eqref{e:optimal_x_1} in \eqref{e:capacity_1}, we obtain the maximum channel capacity as
\begin{equation} \label{e:optimal_C_1}
    \mathcal{C} = \frac{5 \pi (2 \alpha)^{\frac{1}{5}}}{8 e \ln{2}}.
\end{equation}

\subsubsection{Optimal Design Under Low Channel Attenuation}

Next, we consider the case where $x \in \left[f_{p1}^{\text{LED}}, f_{p2}^{\text{LED}} \right)$. The critical point occurs when $x^*=\frac{( 2 \alpha )^{\frac{1}{5}}}{e} = f_{p2}^{\text{LED}}$. Substituting the expression of $\alpha$ at the critical point and solving for $h$ yields
\begin{equation} \label{e:h1}
    h_1 = \frac{2 \sqrt{2 N_0} \pi^3 e^3 C_w L_b}{K^{\text{PA}} \mathcal{R}_L \mathcal{R}_P K^{\text{LNA}} \mu} \left( f_{p2}^{\text{LED}} \right)^{\frac{5}{2}}.
\end{equation}

When the channel attenuation falls below this threshold in \eqref{e:h1}, the channel capacity $\mathcal{C}$ reaches its maximum value for $y^* = f_{p2}^{\text{LED}}$ (see Appendix \ref{sec:AppC}). When $x$ approaches $f_{p2}^{\text{LED}}$, $y^* = f_{p2}^{\text{LED}}$ is aligned with the case in the last subsection where $x^*=y^*$. However, it should be emphasized that the solution in \eqref{e:optimal_x_1}-\eqref{e:optimal_C_1} fail to cover is $x \ll f_{p2}^{\text{LED}}$. Therefore, we shall focus on this particular case in the following.

For $x \ll f_{p2}^{\text{LED}}$, we substitute $y^* = f_{p2}^{\text{LED}}$ into \eqref{e:capacity} and obtain
\begin{equation} \label{e:capacity_2}
    \mathcal{C} \approx \frac{\pi}{4} x \log_2{\left( \frac{\alpha}{\left( f_{p2}^{\text{LED}} \right)^2} \frac{1}{x^3} + 1 \right)}.
\end{equation}
Following a similar approach to the derivation of \eqref{e:optimal_x_1}, the maximum channel capacity is achieved when the optimal poles of the equalizer are set as (see the proof in Appendix \ref{sec:AppD})
\begin{equation} \label{e:optimal_x_2}
    \hspace{-0.2 cm}\left( f_{p1}^{\text{Eq}} \right)^* = x^* = \frac{\alpha^{\frac{1}{3}}}{e \left( f_{p2}^{\text{LED}} \right)^{\frac{2}{3}}}, \quad  \left( f_{p2}^{\text{Eq}} \right)^* = y^* = \left( f_{p2}^{\text{LED}} \right),
\end{equation}
The optimal equalizer parameters are given by 
\begin{eqnarray} 
    R_1^* & = & \frac{e R_g(r_L + 1)}{2 C_w r_L \alpha^{\frac{1}{3}} - 2 e (r_L + 1)},  \\
    L_e^* & = & \frac{e R_g C_w r_L}{2 C_w r_L \alpha^{\frac{1}{3}} - 2 e (r_L + 1)},  \\
    R_2^* & = & 0. \label{e:R2}
\end{eqnarray}
Substituting \eqref{e:optimal_x_2} in \eqref{e:capacity_2}, we obtain the maximum channel capacity as
\begin{equation} \label{e:optimal_C_2}
    \mathcal{C} = \frac{\pi}{4} \frac{\alpha^{\frac{1}{3}}}{e \left( f_{p2}^{\text{LED}} \right)^{\frac{2}{3}}} \frac{3}{\ln{2}}.
\end{equation}

It should be noted that $R_2^*=0$ in \eqref{e:R2} implies that the capacitor $C_e$ in Fig.~\ref{fig:equalizer} is effectively short-circuited, causing the equalizer circuit to degenerate into a first-order design. This observation suggests that when the channel attenuation is severe, the system cannot support a wider bandwidth, and thus a first-order equalizer is sufficient to achieve maximum channel capacity. As the channel conditions further deteriorate, i.e., $ \frac{\alpha ^ {\frac{1}{3}}}{e (f_{p2}^{\text{LED}} )^{\frac{2}{3}}} < f_{p1}^{\text{LED}}$, the optimal poles of the equalizer are set as $x^* = f_{p1}^{\text{LED}}$ and $y^* = f_{p2}^{\text{LED}}$. Substituting the expression for $\alpha$ from \eqref{e:capacity_0} into \eqref{e:optimal_x_2} yields the corresponding channel attenuation as 
\begin{eqnarray} 
    h_2 = \frac{4 \pi^3 e^2 \sqrt{N_0} C_w L_b}{K^{\text{PA}} \mathcal{R}_L \mathcal{R}_P K^{\text{LNA}} \mu} \left( f_{p1}^{\text{LED}} \right)^{\frac{3}{2}}  f_{p2}^{\text{LED}}.
\end{eqnarray}
This implies that the equalizer’s pole coincides with its corresponding zero, rendering bandwidth extension ineffective. The corresponding circuit parameters are $R_1^* = \infty$ and $R_2^*=0$, implying that the signal source and PA in Fig.~\ref{fig:optical_link} are directly connected, bypassing the equalizer. Substituting $x^* = f_{p1}^{\text{LED}}$ and $y^* = f_{p2}^{\text{LED}}$ into \eqref{e:capacity}, we obtain the resulting channel capacity as
\begin{eqnarray} 
    \mathcal{C} = \frac{\pi}{4} \frac{f_{p1}^{\text{LED}} f_{p2}^{\text{LED}} }{f_{p1}^{\text{LED}} + f_{p2}^{\text{LED}}} \log_2{\left( \alpha \frac{f_{p1}^{\text{LED}} + f_{p2}^{\text{LED}}}{\left( f_{p1}^{\text{LED}} \right)^3 \left( f_{p2}^{\text{LED}} \right)^3 } + 1 \right)}.
\end{eqnarray}

In the above, we discussed the optimal system configurations under different channel conditions and provided closed-form expressions for each case. The boundaries for different system settings are defined by $h_1$ and $h_2$, respectively. In summary, the optimal values of $f_{p1}^{\text{Eq}}$ and $f_{p2}^{\text{Eq}}$ are given respectively as
\begin{eqnarray} \label{e:optimal_p1}
    \left( f_{p1}^{\text{Eq}}\right)^* &=& \begin{cases}
        f_{p1}^{\text{LED}}, & h \leq h_2 \\
        \frac{\alpha^{\frac{1}{3}}}{e \left( f_{p2}^{\text{LED}} \right)^{\frac{2}{3}}}, & h_2  < h \ll h_1\\
        \frac{(2 \alpha)^{\frac{1}{5}}}{e}, & \text{otherwise}        
    \end{cases}, \\
    \label{e:optimal_p2}
    \left( f_{p2}^{\text{Eq}}\right)^* &=& \begin{cases}
        f_{p2}^{\text{LED}}, & h \ll h_1 \\
        \frac{(2 \alpha)^{\frac{1}{5}}}{e}, & \text{otherwise}        
    \end{cases}.
\end{eqnarray}
The maximum channel capacity is given by
\begin{equation} \label{e:optimal_C}
    \mathcal{C}^* = \begin{cases}
    \frac{\pi}{4} \frac{f_{p1}^{\text{LED}} f_{p2}^{\text{LED}} }{f_{p1}^{\text{LED}} + f_{p2}^{\text{LED}}} \log_2{\left( \alpha \frac{f_{p1}^{\text{LED}} + f_{p2}^{\text{LED}}}{\left( f_{p1}^{\text{LED}} \right)^3 \left( f_{p2}^{\text{LED}} \right)^3 } + 1 \right)}, & \!\!\!\!h \leq h_2 \\
    \frac{3\pi}{4\ln{2}} \frac{\alpha^{\frac{1}{3}}}{e \left( f_{p2}^{\text{LED}} \right)^{\frac{2}{3}}} , & \!\!\!\!\!\!\!\!\!\!\!\!\!\!\!\!\!\!\!\!\!\!\!\! h_2  < h \ll h_1\\
    \frac{5 \pi }{8 \ln{2}} \frac{(2 \alpha)^{\frac{1}{5}}}{e }, & \!\!\!\!\!\!\!\!\!\!\!\!\!\!\!\!\!\!\!\!\!\!\!\!\text{otherwise}        
    \end{cases}\!\!\!.
\end{equation}

\section{Numerical Results and Discussions}
\label{SecV}
In this section, we present numerical results that evaluate the performance of the optimized VLC system, validating the theoretical framework developed in the previous sections. The parameters used in the numerical simulations are listed in Tab.~\ref{tab:para}, where the equivalent circuit parameters of the LED are selected based on the data for the blue LED in \cite{zhang2024design}.

\begin{table}[!tb]
\centering
\caption{Typical values of system parameters.}
\begin{tabular}{ccc}
     \toprule
     \textbf{Symbol}   & \textbf{Description} & \textbf{Value} \\
     \midrule
     $R_g$ & Standard port impedance   & 50 $\Omega$ \\
     $R_s$ & LED series resistance & 1 $\Omega$ \\
     $r_L$ & LED internal resistance & 0.5 $\Omega$ \\
     $C_w$ & Junction capacitance & 10.8 nF \\
     $L_b$ & Bonding inductance & 28.6 nH \\
     $K^{\text{PA}}$ & Gain of power amplifier & 30 dB \\
     $K^{\text{LNA}}$ & Gain of low-noise amplifier & 30 dB \\
     $N_0$ & Noise power spectral density & -50 dBm \\
     $\mathcal{R}_L$ & Responsivities of LED & 1 \\
     $\mathcal{R}_P$ & Responsivity of photodiode & 1 \\
     $h$ & Channel attenuation & 0.5 \\
     \bottomrule
\end{tabular}
\label{tab:para}
\end{table}

\begin{figure}[!tb]
    \centering
    \includegraphics[width=0.8\linewidth]{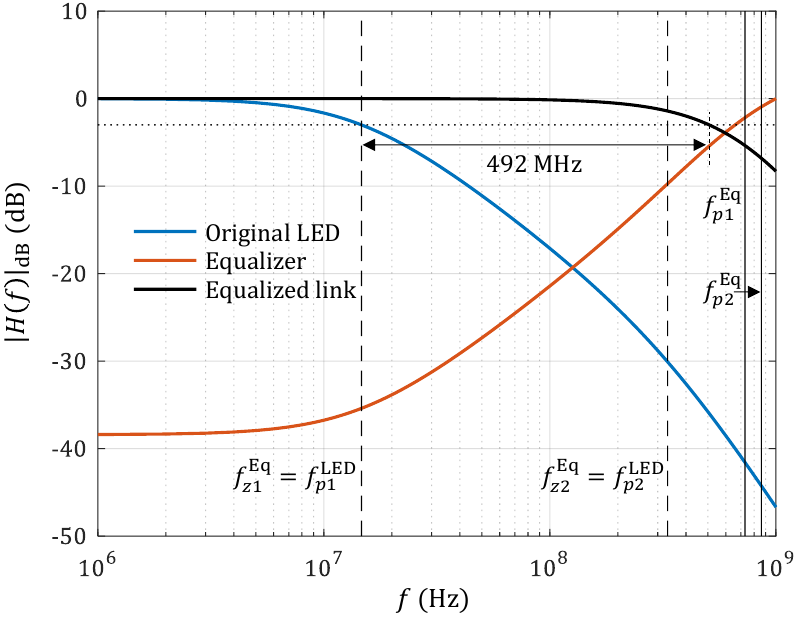}
    \caption{The frequency response of the LED equivalent circuit, the equalizer circuit, and the equalized optical link, illustrating how the equalizer compensates for the LED’s poles and shifts the poles of the cascaded link to higher frequencies. }
    \label{fig:zeros_poles}
\end{figure}

Based on \eqref{e:S21_LED}, \eqref{e:S21_Eq}, and \eqref{e:H}, Fig.~\ref{fig:zeros_poles} presents the normalized frequency response curves, illustrating how the equalizer compensates for the poles introduced by the LED and shifts the poles of the cascaded optical link toward higher frequencies. The response in decibels can be expressed as $|H(f)|_{dB} = 20 \log_{10}{|H(f)|}$.

\begin{figure}
    \centering
    \includegraphics[width=0.8\linewidth]{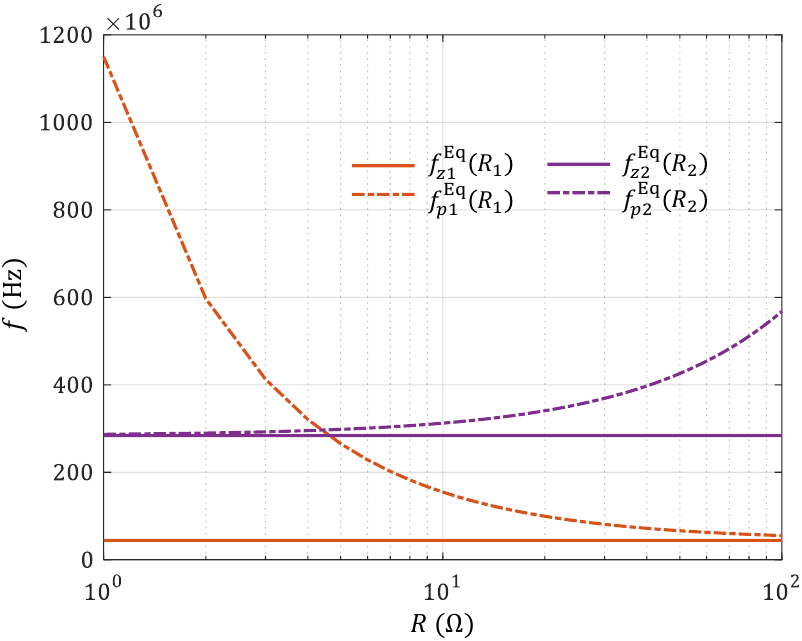}
    \caption{The relationship between the equalizer’s poles and zeros and the circuit parameters. }
    \label{fig:ZPM}
\end{figure}

As demonstrated in \eqref{e:L_e_C_e}, parameters $R_1$ and $R_2$ serve as equivalent representations of the inductive and capacitive components $L_e$ and $C_e$, respectively. Therefore, the corner frequencies of the equalizer shown in \eqref{e:S21_Eq} are directly determined by the values of $R_1$ and $R_2$ as illustrated in Fig.~\ref{fig:ZPM}. The poles of the equalizer circuit are consistently greater than their corresponding zeros, ensuring stable signal enhancement. where different pole placements lead to varying bandwidth-gain trade-offs.

\begin{figure}[!t]
\centering
    \begin{subfigure}[!t]{1\linewidth}
        \centering
        \includegraphics[width=0.8\linewidth]{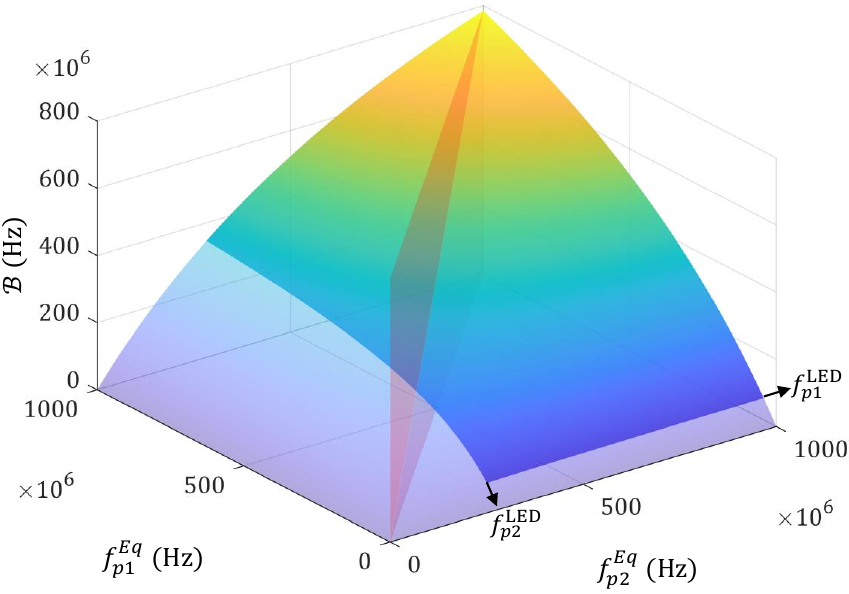}
        \caption{}
    \end{subfigure}
    \begin{subfigure}[!t]{1\linewidth}
        \centering
        \includegraphics[width=0.8\linewidth]{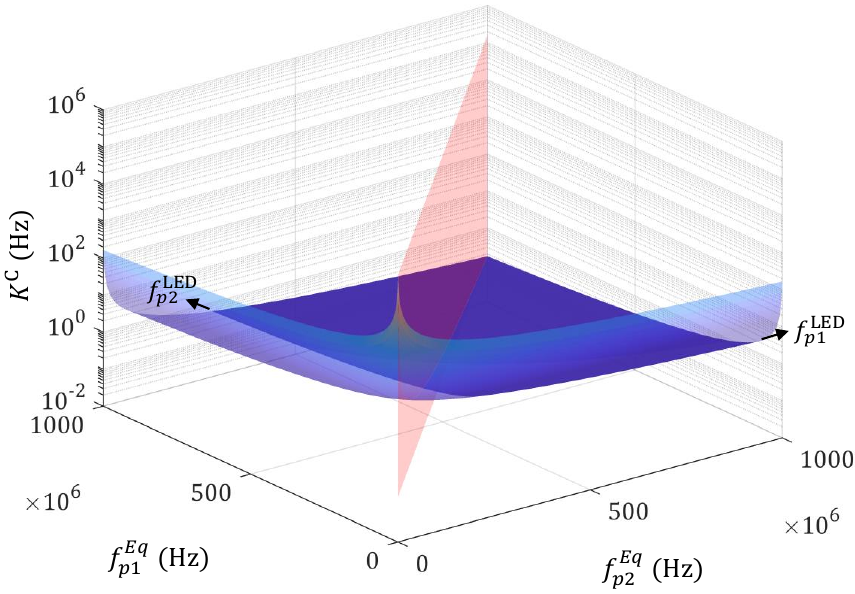}
        \caption{}
    \end{subfigure}
\caption{(a) System bandwidth $\mathcal{B}$ versus link poles. (b) End-to-end channel coefficient $K^{\text{C}}$ of the equalized system versus link poles.}
\label{fig:K_poles}
\end{figure}

To investigate the impact of pole placement on end-to-end channel coefficient $K^{\text{C}}$ in \eqref{e:K^C} and bandwidth $\mathcal{B}$ in \eqref{e:B}, we analyze the variations in system performance under different equalizer pole distributions in Fig.~\ref{fig:K_poles}. Transparency is used to distinguish between symmetric regions for easier observation and the asymmetric regions encountered in practice, specifically $f_{p1}^{\text{Eq}} \in \left[f_{p1}^{\text{LED}}, \infty\right)$ and $f_{p2}^{\text{Eq}} \in \left[f_{p2}^{\text{LED}}, \infty \right)$. As the equalizer pole values increase, the bandwidth consistently expands. However, this expansion is accompanied by a reduction in channel coefficient, highlighting the inherent trade-off between these two parameters.

\begin{figure}
    \centering
    \includegraphics[width=0.8\linewidth]{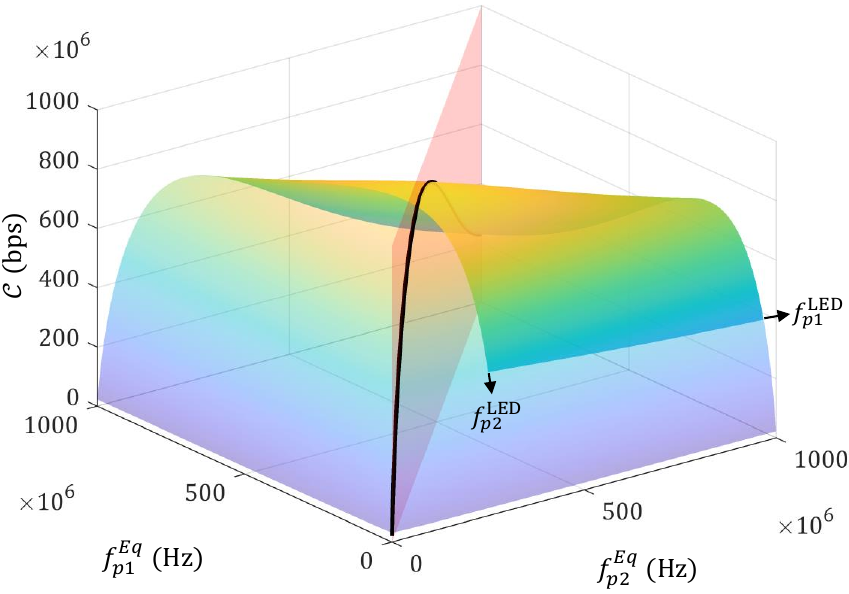}
    \caption{Equalized VLC system capacity $\mathcal{C}$ versus the variation of link poles.}
    \label{fig:capacity}
\end{figure}

In Fig.~\ref{fig:capacity}, based on \eqref{e:capacity}, we present the channel capacity as a function of the optimal pole values. It is observed that increasing poles of the equalizer circuit leads to a rise in channel capacity at low frequencies, primarily due to bandwidth expansion, while at high frequencies, the capacity declines as reduced channel coefficient becomes the dominant factor.

\begin{figure}
    \centering
    \includegraphics[width=0.8\linewidth]{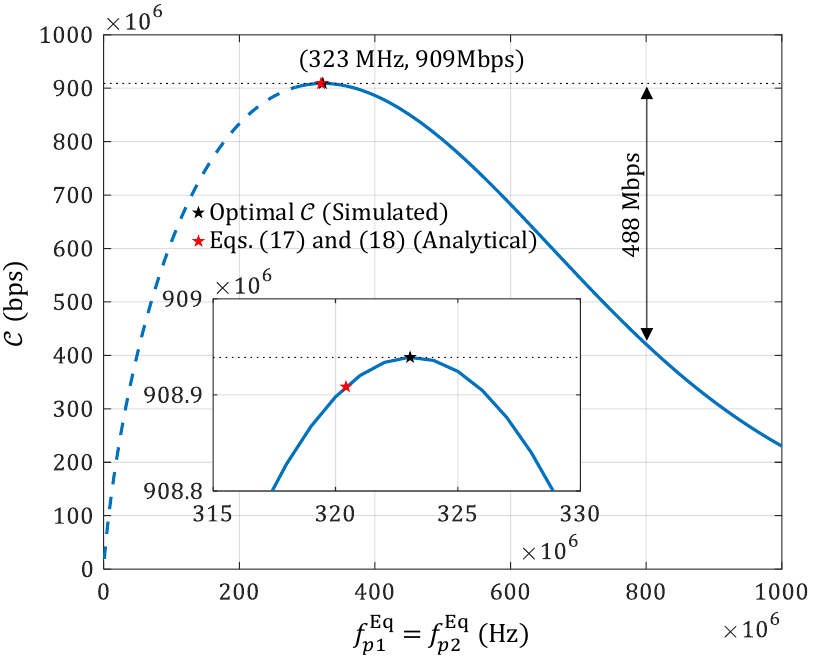}
    \caption{The channel capacity for large values of channel attenuation.}
    \label{fig:capacity_lemma4}
\end{figure}

When the optimal values of the two poles lie within the same region, the constraints obtained under large values of channel attenuation enable us to simplify the analysis by focusing solely on the curve along the $x=y$ plane as shown in Fig.~\ref{fig:capacity_lemma4}. The numerical results for the maximum channel capacity are presented in this figure, showing strong agreement with the analytical values derived in \eqref{e:optimal_x_1} and \eqref{e:optimal_C_1}, as evidenced by the near overlap of their corresponding data points. Furthermore, the results confirm that, for a channel attenuation of 0.5, the optimal bandwidth enhances the data rate by 488 Mbps compared to an 800 MHz optical link across typical VLC scenarios.

\begin{figure}[!t]
\centering
    \begin{subfigure}[!t]{1\linewidth}
        \centering
        \includegraphics[width=0.8\linewidth]{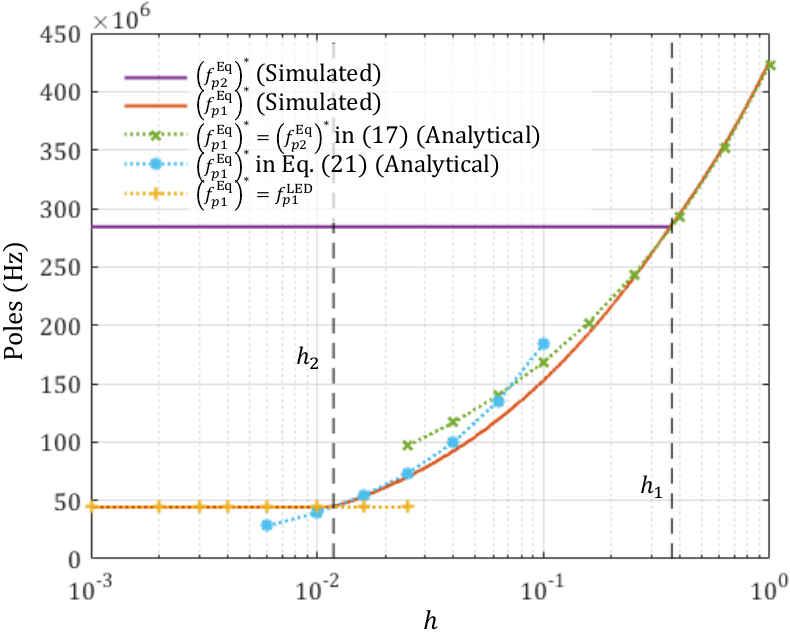}
        \caption{}
    \end{subfigure}
    \begin{subfigure}[!t]{1\linewidth}
        \centering
        \includegraphics[width=0.8\linewidth]{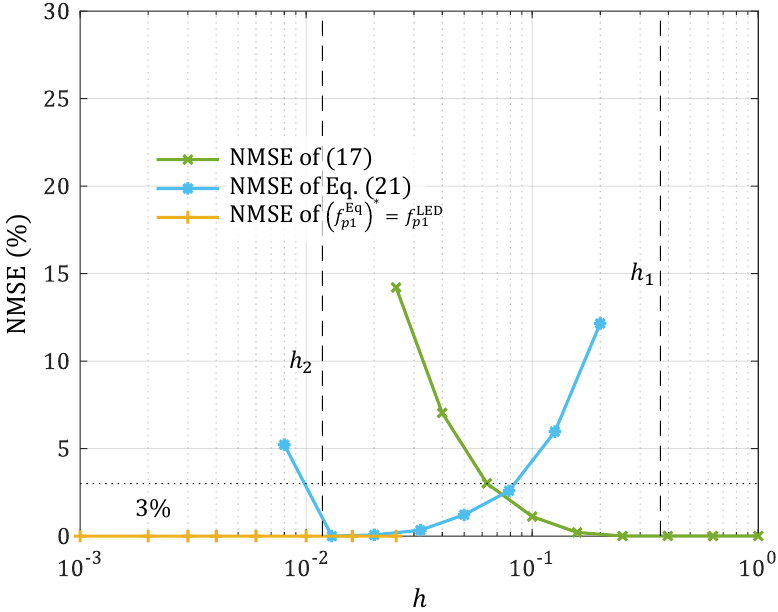}
        \caption{}
    \end{subfigure}
\caption{(a) The relationship between the optimal poles and the variation of the channel attenuation $h$. (b) The NMSE of the analytically derived optimal poles.}
\label{fig:poles_h}
\end{figure}

As channel attenuation changes, the optimal pole values obtained in \eqref{e:optimal_p1} and \eqref{e:optimal_p2} adapt accordingly. The relationship between pole values under different conditions is shown in Fig.~\ref{fig:poles_h} (a), with the normalized mean square error (NMSE) of the analytical solutions shown in Fig.~\ref{fig:poles_h} (b). Under better channel conditions, the optimal values of the two poles can be large and fall within the same feasible region of both poles, where their optimal values are equal. When the channel conditions worsen, pole $f_{p2}^{\text{Eq}}$ can only take the minimum value within the acceptable range, while pole $f_{p1}^{\text{Eq}}$ decreases as the channel attenuation decreases. Therefore, in practical deployment, a switchable equalizer structure, e.g., such as proposed in \cite{kisacik2022130}, can be considered to accommodate optimal system configurations under varying channel attenuation conditions. The NMSE of analytical values shows that throughout the entire channel variation process, the fitting error of the analytical values remains within 3\%.

\begin{figure}[!t]
\centering
    \begin{subfigure}[!t]{1\linewidth}
        \centering
        \includegraphics[width=0.8\linewidth]{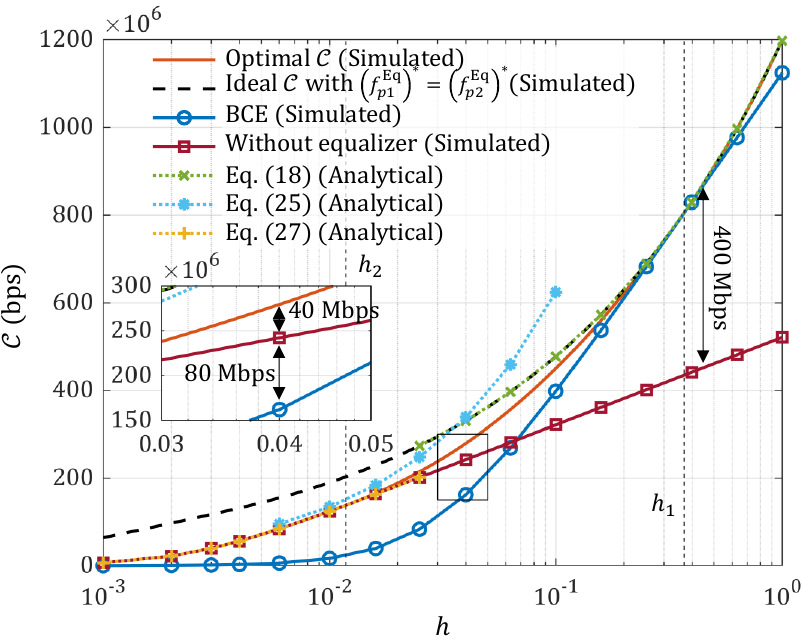}
        \caption{}
    \end{subfigure}
    \begin{subfigure}[!t]{1\linewidth}
        \centering
        \includegraphics[width=0.8\linewidth]{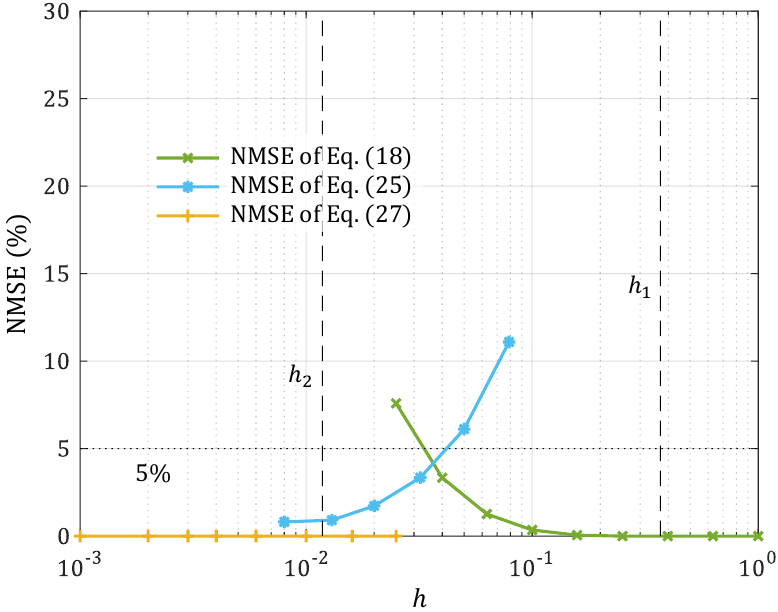}
        \caption{}
    \end{subfigure}
\caption{(a) Maximum channel capacity versus the channel attenuation $h$. (b) The NMSE of the analytically derived maximum channel capacities.}
\label{fig:capacity_h}
\end{figure}

The numerical channel capacity and the analytical expressions for the channel capacity under different conditions are plotted in Fig.~\ref{fig:capacity_h} (a), with the NMSE of the analytical solutions shown in Fig.~\ref{fig:capacity_h} (b) remaining within 5\%. It is observed that the achievable channel capacity increases as the channel attenuation increases. As shown in Tab.~\ref{tab:review}, although some previous works have considered second-order equalizer circuits, they are typically based on fixed channel conditions. However, when the channel varies, a single static system configuration can no longer ensure optimal performance. To clearly illustrate the benefits of the proposed capacity-based optimization approach, we consider two benchmark schemes:
\begin{itemize}
    \item \textbf{ Bandwidth-Centric Equalizer (BCE)}: $f_{p1}^{\text{Eq}} = f_{p2}^{\text{Eq}} = f_{p2}^{\text{LED}}$, , representing the critical configuration for $h = h_1$ that enables bandwidth extension into the hundreds of megahertz range. 
    \item \textbf{ Without equalizer}: $f_{p1}^{\text{Eq}} = f_{p1}^{\text{LED}}$, and  $f_{p2}^{\text{Eq}} = f_{p2}^{\text{LED}}$, representing the critical configuration for $h = h_2$, resulting in a bandwidth limited to just a few tens of megahertz.
\end{itemize}
Our results show that the proposed capacity-centric equalizer (CCE) consistently achieves the best performance across all channel attenuations. Under favorable channel conditions, $h=0.4$, the capacity gap between the system equipped with CCE and that without an equalizer reaches up to 400 Mbps, and the BCE also provides a notable capacity gain. As the channel deteriorates, the CCE still maintains a gain of approximately 40 Mbps at $h=0.04$, whereas the severe signal power attenuation introduced by the BCE leads to a rapid decline in capacity falling even below that of the system without an equalizer. Therefore, while blindly extending the bandwidth does not always result in capacity gains, a CCE tailored to the specific channel attenuation can significantly enhance overall capacity.

\begin{figure}
    \centering
    \includegraphics[width=0.8\linewidth]{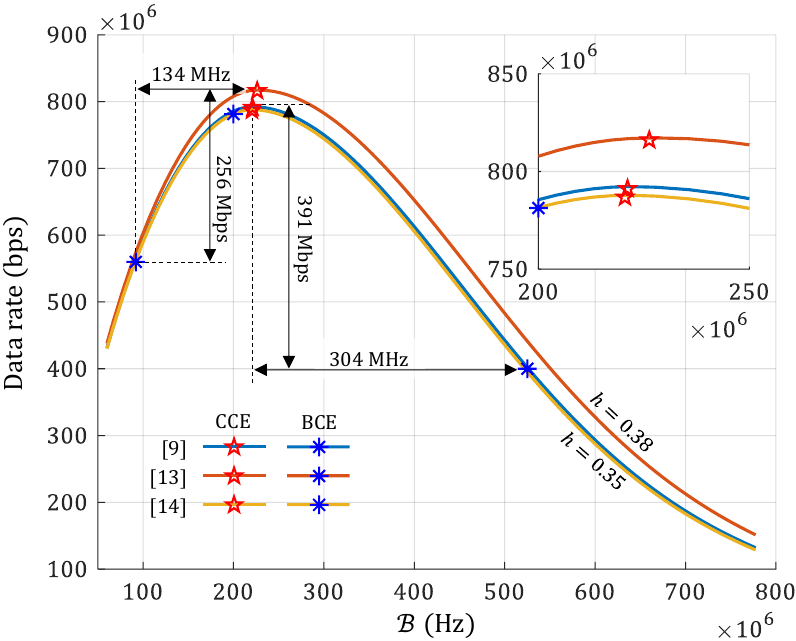}
    \caption{Comparison of proposed CCE with bandwidth-optimized equalizers under different channel attenuations. }
    \label{fig:comparison}
\end{figure}

In Fig.~\ref{fig:comparison}, we plot curves illustrating the relationship between channel capacity and bandwidth under different channel  conditions. These curves are generated by combining the specific expressions for bandwidth and channel capacity provided in \eqref{e:B} and \eqref{e:optimal_C}, using the equalized link poles derived in \eqref{e:optimal_p1} and \eqref{e:optimal_p2} as intermediate variables. We further compare these results with several existing  VLC systems, which generally share similar model parameters as listed in Tab.~\ref{tab:para}. For each system, both the reported test points and the maximum achievable rates predicted by the CCE model are plotted. As shown in Fig.~\ref{fig:comparison}, these systems exhibit similar channel attenuation characteristics, and their optimal bandwidths are generally found to cluster around 225 MHz. In particular:
\begin{itemize}
    \item In \cite{zhang2024design}, a low-complexity analog equalizer was used to achieve a flat bandwidth exceeding 500 MHz - over 300 MHz wider than that of the CCE. However, when the channel attenuation is insufficient to support higher data rates, blindly expanding the modulation bandwidth severely degrades the signal power, ultimately resulting in a 391 Mbps reduction in data rate.
    \item In \cite{chen2021digital}, digital equalization was employed to achieve higher data rates. However, without physical-layer optimization for capacity, the resulting bandwidth was 134 MHz narrower than that of the CCE. Although this preserved signal power, the nonlinear trade-off between bandwidth and gain still led to a 256 Mbps capacity loss.
    \item The bandwidth settings adopted in \cite{li2023performance} closely match the optimal values predicted by the CCE model, resulting in significantly higher data rates compared to systems with either excessively wide or narrow bandwidths.
\end{itemize}

This comparison with benchmark systems further demonstrates the superiority of the proposed capacity-optimized design over traditional bandwidth-centric equalization approaches.

\section{Conclusion}
\label{SecVI}

Analog pre-equalization is a commonly used technique in VLC systems to extend modulation bandwidth of LEDs. In this paper, we introduced a systematic design methodology for analog pre-equalizers that optimally balance bandwidth and frequency response to maximize channel capacity. We observed that in good channel conditions, the signal power attenuation caused by a wider modulation bandwidth becomes less significant, and the broader modulation bandwidth can support a higher channel capacity. Conversely, in poor channel conditions, a first-order equalizer is sufficient to achieve the optimal system configuration. We also derived closed-form expressions for the optimal equalizer parameters and the corresponding maximum achievable capacity, and validated them through extensive simulations. These results provided a theoretical foundation for the design of VLC systems with bandwidth-limited LED transmitters that maximize channel capacity under various scenarios and channel conditions. 

\appendices
\section{Proof of \eqref{e:B}}\label{sec:AppA}
In this appendix, we derive the 3-dB bandwidth of the equalized system given in (10). To approximate it, we use the concept of noise equivalent bandwidth, which represents the bandwidth of an ideal filter that would generate the same noise power as the actual filter. For optical links with sharp cutoff characteristics, the noise equivalent bandwidth provides an accurate estimate of the 3-dB bandwidth \cite{carlson2010communication}. It is mathematically defined as
\begin{equation}
    \mathcal{B} = \frac{1}{\left| H(f) \right|_{\max}^2} \int_0^{\infty} {\left| H(f) \right|^2 df}.
\end{equation}
Substituting the expression of $H(f)$ given in (9), the 3dB bandwidth of the equalized VLC system can be calculated as
\begin{eqnarray}
\mathcal{B} 
&\overset{(a)}{=}& {\frac{1}{\left(K^{\text{PA}} K^{\text{Eq}} K^{\text{LED}}\right)^2} 
\int_0^{\infty} \left| \frac{K^{\text{PA}} K^{\text{Eq}} K^{\text{LED}}}
{\left( \frac{j f}{f_{p1}^{\text{Eq}}} + 1 \right)\left( \frac{j f}{f_{p2}^{\text{Eq}}} + 1 \right)} \right|^2 df} \notag\\
&=& \int_0^{\infty} \frac{1}{\left( \frac{f}{f_{p2}^{\text{Eq}}} + \frac{f}{f_{p1}^{\text{Eq}}} \right)^2 
+ \left( \frac{f}{f_{p1}^{\text{Eq}}} \frac{f}{f_{p2}^{\text{Eq}}} - 1 \right)^2} \, df \notag\\
&\overset{(b)}{=}& {\int_0^{\infty} \frac{1}{\left( \left( \frac{f}{f_{p1}^{\text{Eq}}} \right)^2 + 1 \right)
\left( \left( \frac{f}{f_{p2}^{\text{Eq}}} \right)^2 + 1 \right)} \, df} \notag\\
&\overset{(c)}{=}& \frac{(f_{p2}^{\text{Eq}})^2}{(f_{p2}^{\text{Eq}})^2 - (f_{p1}^{\text{Eq}})^2} 
\int_0^{\frac{\pi}{2}} \frac{1}{\tan^2 u + 1} \cdot \frac{f_{p1}^{\text{Eq}}}{\cos^2 u} \, du \notag\\
&& - \frac{(f_{p1}^{\text{Eq}} )^2}{(f_{p2}^{\text{Eq}})^2 - (f_{p1}^{\text{Eq}})^2} 
\int_0^{\frac{\pi}{2}} \frac{1}{\tan^2 v + 1} \cdot \frac{f_{p2}^{\text{Eq}}}{\cos^2 v} \, dv \notag\\
&=& \frac{\pi}{2} \frac{f_{p1}^{\text{Eq}} f_{p2}^{\text{Eq}}}{f_{p1}^{\text{Eq}} + f_{p2}^{\text{Eq}}},
\end{eqnarray}
where (a) follows from (24), (b) follows from $f_{p1}^{\text{Eq}} = \frac{2R_1+R_g}{4\pi L_e}$ and $f_{p2}^{\text{Eq}} = \frac{2R_g+R_2}{2R_g C_e R_2}$, and (c) follows from the variable substitution where $\frac{f}{f_{p1}^{\text{Eq}}} = \tan{u}$ and $\frac{f}{f_{p2}^{\text{Eq}}} = \tan{v}$, along with $\frac{1}{f_{p1}^{\text{Eq}}} df = \frac{1}{\cos^2{u}} du $ and $\frac{1}{f_{p2}^{\text{Eq}}} df = \frac{1}{\cos^2{v}} dv $. Therefore, (10) is proved.

\section{Proof of \eqref{e:optimal_x_1}}\label{sec:AppB}
In this appendix, we present a detailed derivation on how the solution in (25) is obtained. We begin by recalling (24) as
\begin{eqnarray} 
    \log_2{\left(  \frac{2 \alpha}{x^5} + 1 \right)} = \frac{5}{\frac{x^5}{\alpha} + 1}.
\end{eqnarray}
Setting $ \frac{2\alpha}{x^5} +1 = u $ with $ u > 1 $, yields $ x = \left( \frac{2\alpha}{u - 1} \right)^{\frac{1}{5}} $ and
\begin{eqnarray}
&& \hspace{0.8 cm} \ln{u} = \frac{5(u - 1)}{u} = 5\left(1 - \frac{1}{u}\right) \\
\Rrightarrow && \hspace{1.53 cm}  (\ln{u} - 5)u = -5 \notag\\
\Rrightarrow && \hspace{-0.5 cm} (\ln{u} - 5) \cdot \exp(\ln{u} - 5) = -\frac{5}{e^5} \notag
\end{eqnarray}
By setting $ \ln u - 5 = t $ with $ t > -5 $, we get
\begin{eqnarray}
t \exp(t) = -\frac{5}{e^5}
\end{eqnarray}
To solve this equation, we define a new function $ F_t(t) = t \exp(t) $. It is evident that $ F_t(t = 0) = F_t( t = -\infty) = 0 $. Furthermore, note that $F_t'(t) = \exp(t) + t \exp(t)$, which implies that the minimum value of $ F_t(t) $ occurs at $ t = -1 $, i.e., $ F_{t,\min}(t) = F_t(-1) = -\frac{1}{e} $. Since $-\frac{1}{e} < -\frac{5}{e^5} < 0$, the function $ F_t(t) = -\frac{5}{e^5} $ has one root in each of the intervals $ (-\infty, -1) $ and $ (-1, 0) $. Clearly, the root in the interval $ (-\infty, -1) $ is $ t_1 = -5 $. In the interval $ (-1, 0) $, a Taylor expansion yields a first-order approximation of $ F_t(t) \approx t $. Consequently, the other root is $ t_2 = -\frac{5}{e^5} $. Since $ t > -5 $, the unique solution for $ t $ is $t = -\frac{5}{e^5}$.

Recalling that $ u = \exp(t + 5) $, the solution for $ u $ is given by
\begin{eqnarray}
u = \exp\left(-\frac{5}{e^5} + 5\right).
\end{eqnarray}
Additionally, since $ x = \left( \frac{2\alpha}{u - 1} \right)^{\frac{1}{5}} $, the only solution for $ x $ is
\begin{eqnarray}
x^* = \left( \frac{2\alpha}{e^5 - 1} \right)^{\frac{1}{5}} \approx \frac{(2\alpha)^{\frac{1}{5}}}{e}, 
\end{eqnarray}
where the approximation is due to $ -\frac{5}{e^5} \ll 5 $, and $ e^5 \gg 1 $. Therefore, (25) is proved.

\section{Proof of Optimal Conditions under Low Channel Attenuation}\label{sec:AppC}
In this appendix, we prove that the channel capacity $\mathcal{C}$ reaches its maximum at $y^* = f_{p2}^{\mathrm{LED}}$. For this proof, it suffices to demonstrate that $\frac{\partial C}{\partial y} < 0 \quad \text{for} \quad y \in \left[ f_{p2}^{\mathrm{LED}}, \infty \right)$, meaning that 
\begin{eqnarray}
\frac{\partial \mathcal{C}}{\partial y} &=& \frac{\pi}{4}   \frac{x^2}{(x + y)^2} \log_2\left(1 + \alpha   \frac{x + y}{x^3 y^3} \right) \\
&&\hspace{1.5 cm} + \frac{\pi}{4}   \frac{\alpha}{\ln 2}   \frac{x}{x + y}   \frac{-2y - 3x}{x^3 y^3 + \alpha(x + y)} \notag\\
&=& \frac{\pi}{4}   \frac{x}{x + y} \left[ \frac{x}{x + y} \log_2\left(1 + \alpha   \frac{x + y}{x^3 y^3} \right) \right.  \notag\\
&&\hspace{1.7 cm} + \left. \frac{\alpha}{\ln 2}   \frac{-2y - 3x}{x^3 y^3 + \alpha(x + y)} \right] < 0.\notag
\end{eqnarray}
Since $\frac{x}{x + y} > 0$, it becomes sufficient to prove that
\begin{equation}
\frac{x}{x + y}  \! \log_2\left(1  \! + \!  \alpha   \frac{x + y}{x^3 y^3} \right) \! + \! \frac{\alpha}{\ln 2}   \frac{-2y - 3x}{x^3 y^3  \! + \!  \alpha(x + y)} \!  < \!  0,
\end{equation}
or
\begin{equation}
\log_2\left(1 + \alpha   \frac{x + y}{x^3 y^3} \right) < \frac{\alpha}{\ln 2}   \frac{2y + 3x}{x^3 y^3 + \alpha(x + y)}   \frac{x + y}{x}.
\end{equation}
By setting
\begin{equation}
\hspace{-0.2 cm} F_y(y) = \log_2  \! \left( \! 1  \! + \!  \alpha   \frac{x + y}{x^3 y^3} \!  \right) \!  - \!  \frac{\alpha}{\ln 2}   \!  \frac{2y + 3x}{x^3 y^3  \! +  \! \alpha(x + y)}  \!   \frac{x + y}{x},
\end{equation}
it suffices to prove that: 
1) $F_y(\beta) < 0$ holds for $x \in [f_{p1}^{\text{LED}}, \beta)$;
2) $F_y'(y) < 0$ holds for $y \in [\beta, \infty)$, where we set $\beta \triangleq f_{p2}^{\text{LED}}$ for brevity.

First, we prove the first point. By taking the partial derivative of $F_y(\beta)$ with respect to $x$, we obtain

\begin{eqnarray}
&&\hspace{-0.8 cm }\frac{\partial F_y(\beta)}{\partial x} \!=\! \frac{1}{\ln 2}   \frac{\alpha}{1 \!+\! \alpha \frac{x + \beta}{x^3 \beta^3}}      \frac{x^3 \beta^3 \!-\! (x + \beta) 3x^2 \beta^3}{x^6 \beta^6} \!-\! \frac{\alpha}{\ln 2}  \notag\\
&&\hspace{0 cm } \cdot \frac{\begin{bmatrix}
    (3(x + \beta) + (2\beta + 3x))(x^3 \beta^3 + \alpha(x + \beta))x \\
    - (2\beta \!+\! 3x)\!(x \!+\! \beta)\!(x^3 \beta^3 \!+\! \alpha(x \!+\! \beta) \!+\! x(3x^2 \beta^3 \!+\! \alpha))
\end{bmatrix}}{(x^3 \beta^3 + \alpha(x + \beta))^2 x^2}  \notag \\
&&\hspace{0.4 cm } = \frac{\alpha}{\ln 2}   \frac{1}{(x^3 \beta^3 + \alpha(x + \beta))}   \frac{1}{x}   \frac{1}{\beta^3 (x^3 \beta^3 + \alpha(x + \beta))x} \notag\\
&&\hspace{0.8 cm } \cdot \left[(2\beta + 3x)(x + \beta)x(3x^2 \beta^3 + \alpha) \beta^3 \right. \notag\\
&&\hspace{0.8 cm } \left. + (x^3 \beta^3 + \alpha(x + \beta)) \beta^3 (x(3\beta - x) + 2\beta^2) \right].
\end{eqnarray}
Since $x < \beta$, each term in (54) is positive, leading to $\frac{\partial F_y(\beta)}{\partial x} > 0$.

For $x \in [f_{p1}^{\text{LED}}, \beta)$,
\begin{eqnarray}
F_y(\beta) < F_y(\beta)|_{x=\beta} &=& \log_2 \! \left(1 \! + \! \frac{2\alpha \beta}{\beta^6}\right) \! - \! \frac{\alpha}{\ln 2}   \frac{10\beta}{\beta^6 \! + \! 2\alpha \beta} \notag\\
&=& \log_2{u} - \frac{5}{\ln 2}   \frac{u - 1}{u},
\end{eqnarray}
where (a) follows from setting $1 + \frac{2\alpha}{(f_{p2}^{\text{LED}})^5} = u$, and $u > 0$. 
Since $\alpha < \frac{1}{2}(e f_{p2}^{\text{LED}})^5$ holds for $x \in [f_{p1}^{\text{LED}}, f_{p2}^{\text{LED}})$, we obtain that $u = 1 + \frac{2\alpha}{(f_{p2}^{\text{LED}})^5} < 1 + \frac{(e f_{p2}^{\text{LED}})^5}{(f_{p2}^{\text{LED}})^5} = 1 + e^5$.

Now, define a new function as
\begin{equation}
F_u(u) = \log_2{u} - \frac{5}{\ln 2}   \frac{u - 1}{u}.
\end{equation}
It follows that
\begin{equation}
F_u(1) = 0,
\end{equation}
and
\begin{equation}
F_u'(u) = \frac{1}{\ln 2}   \frac{1}{u} \left(1 - \frac{5}{u}\right).
\end{equation}
Therefore, this function first decreases and then increases, with $u = 5$ as the dividing point. Also, we have
\begin{eqnarray}
F_u(u = 1 + e^5) &=& \log_2(1 + e^5) - \frac{5}{\ln 2}   \frac{e^5}{1 + e^5} \notag\\
&\approx& \log_2(e^5) - \frac{5}{\ln 2} = 0.    
\end{eqnarray}
Thus for $1 < u < 1 + e^5$, we always have $F_u(u) \leq 0$, which implies $F_y(\beta) < 0$.

Next, we prove the second point. Taking the derivative of $F_y(y)$ with respect to $y$, we obtain
\begin{eqnarray}
&&\hspace{-0.5 cm}F_y'(y) = \frac{1}{\ln 2}   \frac{1}{x^3 y^3 + \alpha(x + y)}   \alpha   \frac{x^3 y^3 - (x + y)3x^3 y^2}{x^3 y^3} \notag\\
&&\hspace{0.2 cm} - \frac{\alpha}{\ln 2}   \frac{\begin{bmatrix}
    [2(x + y) + (2y + 3x)](x^3 y^3 + \alpha(x + y))x \\- (2y + 3x)(x + y)(3x^3 y^2 + \alpha)
\end{bmatrix}}{(x^3 y^3 + \alpha(x + y))^2 x^2} \notag\\
&&\hspace{0.5 cm}= -\frac{1}{\ln 2}   \frac{1}{(x^3 y^3 + \alpha(x + y))^2}   \frac{1}{x^2 y} \left[2y^5 x^3 (2x - 3) \right.\notag\\
&&\hspace{0.2 cm}\left.+ 3y^4 x^4 (4x - 5) + y^3 (3x^5(x - 9) + 2\alpha(2x - 1)) \right.\notag\\
&&\hspace{0.2 cm}\left.+ \alpha y^2 x(16x - 5) + 3\alpha y x^2(5x - 1) + 3\alpha x^4\right].
\end{eqnarray}
For $x \in [f_{p1}^{\text{LED}}, \beta)$, where $f_{p1}^{\text{LED}}$ is a parameter on the order of MHz, it is evident that each term in the expression is positive, leading to $F_y'(y) < 0$.

In conclusion, $\frac{\partial \mathcal{C}}{\partial y} < 0$ holds for all $y \in [\beta, \infty)$. Therefore, $\mathcal{C}$ reaches its maximum value at $y^* = \beta = f_{p2}^{\text{LED}}$.

\section{Proof of \eqref{e:optimal_x_2}}\label{sec:AppD}
In this appendix, we derive the optimal poles for the case of small channel attenuation, as given in equation~(33). According to the proof in Appendix \ref{sec:AppC}, we have $y^* = f_{p2}^{\text{LED}}$. Therefore, we obtain
\begin{equation}
\frac{2R_g + R_2}{2R_g C_e R_2} = \frac{1}{C_e R_2}.
\end{equation}
Therefore, the optical $R_2$ is obtained as
\begin{equation}
R_2^* = 0,
\end{equation}
which proves (36). From the circuit structure shown in Fig. 3, $R_2$ shorts $C_e$, causing the equalizer circuit to degrade into a first-order equalizer. This is reasonable because, under poor channel conditions, a second-order equalizer is no longer necessary to achieve equalization over a wide range. Instead, a flat bandwidth within a lower frequency range is sufficient to meet the requirement of maximizing channel capacity. To obtain the maximum of $\mathcal{C}$, we set $\frac{\partial \mathcal{C}}{\partial x} = 0$ and get
\begin{equation}
\frac{\partial \mathcal{C}}{\partial x} = \frac{\pi}{4} \log_2\left( \frac{\alpha}{\beta^2 x^3} + 1 \right)
+ \frac{\pi}{4} x  \frac{1}{\left( \frac{\alpha}{\beta^2 x^3} + 1 \right)}  \frac{1}{\ln 2}  \frac{\alpha}{\beta^2}  \left(-\frac{3}{x^4}\right).
\end{equation}
This solution yields
\begin{equation}
\ln \left( \frac{\alpha}{(f_{p2}^{\text{LED}})^2 x^3} + 1 \right)
= \frac{1}{\frac{\alpha}{(f_{p2}^{\text{LED}})^2 x^3} + 1}  \frac{3\alpha}{(f_{p2}^{\text{LED}})^2 x^3}.
\end{equation}
Setting $\frac{\alpha}{(f_{p2}^{\text{LED}})^2 x^3} + 1 = u> 1$ and using $x = \left( \frac{\alpha}{(f_{p2}^{\text{LED}})^2 (u - 1)} \right)^{\frac{1}{3}}$, we have
Substituting into the equation gives
\begin{equation}
\ln{u} = \frac{1}{u}  \frac{3\alpha}{(f_{p2}^{\text{LED}})^2}  \frac{1}{\frac{\alpha}{(f_{p2}^{\text{LED}})^2 } \frac{1}{u - 1}},
\end{equation}
which can be simplified as
\begin{equation}
\ln{u} = 3  \frac{u - 1}{u}.
\end{equation}
This is similar to (51). Therefore, following the same steps, we can obtain
\begin{equation}
u^* = \exp\left( -\frac{3}{e^3} + 3 \right) \approx \exp(3),
\end{equation}
and the optimal $x$ is given by
\begin{equation}
x^* = \left( \frac{\alpha}{(f_{p2}^{\text{LED}})^2 } \frac{1}{e^3 - 1} \right)^{\frac{1}{3}} \approx \frac{\alpha^{\frac{1}{3}}}{e (f_{p2}^{\text{LED}})^{\frac{2}{3}}} .
\end{equation}
Therefore, (33) is proved. Then, by combining the relationship between $x^*$ and the equalizer parameters given in (8), we obtain
\begin{eqnarray}
R_1^* &=& \frac{R_g}{\frac{2 f_{p1}^{\text{Eq}}}{f_{p1}^{\text{LED}}} - 2}
= \frac{R_g}{\frac{\frac{2 \alpha^{\frac{1}{3}}}{e \left( \frac{R_s + R_g}{2\pi L_b} \right)^{\frac{2}{3}}}}{ \frac{r_L + 1}{2\pi C_w r_L}}   - 2 } \\
&=& \frac{1}{2}  \frac{R_g e (R_s + R_g)^{\frac{2}{3}} (r_L + 1)}{ \alpha^{\frac{1}{3}} (2\pi L_b)^{\frac{2}{3}} 2\pi C_w r_L - e (R_s + R_g)^{\frac{2}{3}} (r_L + 1)}, \notag
\end{eqnarray}

and
\begin{eqnarray}
L_e^* &=& \frac{R_1^* C_w r_L}{r_L + 1} \\
&=& \frac{1}{2}  \frac{e C_w r_L R_g (R_s + R_g)^{\frac{2}{3}}}{\alpha^{\frac{1}{3}} (2\pi L_b)^{\frac{2}{3}} 2\pi C_w r_L - e (R_s + R_g)^{\frac{2}{3}} (r_L + 1)}, \notag
\end{eqnarray}
proving (34) and (35).

Finally, the maximum channel capacity in (37) can be obtained by substituting $x^*$ into (32), as
\begin{equation}
\mathcal{C} = \frac{\pi}{4} x \log_2\left( \frac{\alpha}{(f_{p2}^{\text{LED}})^2 x^3} + 1 \right)  
\approx \frac{\pi}{4}  \frac{\alpha^{1/3}}{e (f_{p2}^{\text{LED}})^{2/3}}  \frac{3}{\ln 2} .
\end{equation}

\bibliographystyle{IEEEtran}
\bibliography{ref.bib}


\end{document}